\documentclass[aps,pra,twocolumn,superscriptaddress,longbibliography,nofootinbib,floatfix]{revtex4-2}

\usepackage{amsmath,amssymb,amsthm,bm}
\usepackage{graphicx}
\usepackage{booktabs}
\usepackage{microtype}
\usepackage[colorlinks=true,linkcolor=blue,citecolor=blue,urlcolor=blue]{hyperref}

\newtheorem{theorem}{Theorem}
\newtheorem{proposition}{Proposition}

\newtheorem{corollary}{Corollary}
\theoremstyle{remark}

\newcommand{\dd}{\mathrm d}
\newcommand{\R}{\mathbb R}
\newcommand{\kvec}{\bm k}
\newcommand{\qvec}{\bm q}
\newcommand{\G}{\mathcal G}
\newcommand{\Khat}{\widehat K}
\newcommand{\Hhat}{\widehat H}

\begin{document}

\title{Geometry-controlled correlated electric-field noise in enclosed ion traps from billiard return spectra}
\hypersetup{
  pdftitle={Geometry-controlled correlated electric-field noise in enclosed ion traps from billiard return spectra},
  pdfauthor={Ayush Nadiger}
}

\author{Ayush Nadiger}
\email{anadiger@umass.edu}
\affiliation{Department of Electrical and Computer Engineering, University of Massachusetts Amherst, Amherst, Massachusetts 01003, USA}
\affiliation{Department of Mathematics and Statistics, University of Massachusetts Amherst, Amherst, Massachusetts 01003, USA}

\date{April 2026}

\begin{abstract}
We ask how passive conducting geometry determines the spatial structure of electric-field noise seen by trapped ions.  From a boundary-potential covariance and the Dirichlet Green function we construct the $N$-ion electric-field cross-spectral matrix and its blockwise motional Kossakowski generator.  In a parallel slab, billiard unfolding turns the electrostatic response into a return-depth measure and yields an exact return-pair functional for arbitrary stationary surface spectra.  For every finite equal-height ion array, a passive cover obeys the matrix inequalities $C^{(y)}_h-C^{(y)}_\infty\succeq0$ and $C^{(x)}_\infty-C^{(x)}_h\succeq0$: all collective normal-field coordinates acquire more absolute noise, while all collective tangential coordinates acquire less.  At $h=2d$ the local-noise single-ion ratios are exactly $\zeta(3)$ and $\eta(3)$.

Diagonalizing the equal-frequency covariance identifies collective environmental noise eigenchannels and a geometry-dependent noise rank; within a degenerate frequency block these become the Lindblad jump channels.  In a ten-ion example, closing the cover to $h=2d$ lowers the participation rank from $5.61$ to $5.11$ while increasing the leading channel's share from $23.7\%$ to $28.6\%$; a primitive M\o lmer--S\o rensen calculation shows how the projected covariance sets the weak-heating gate exposure.  Beyond parallel walls, specular paths emerge as large-$q_z$ saddles of the screened boundary operator.  Across 16 curved covers, the fitted electrostatic decay exponent correlates at $0.9991$ with the independently computed shortest specular excess length, while separate tests resolve focusing and competing saddles.
\end{abstract}

\maketitle

\section{Introduction}

Electric-field noise near conducting surfaces limits motional coherence and entangling-gate performance in trapped-ion processors \cite{Turchette2000,Deslauriers2006,Brownnutt2015,Sedlacek2018,Webb2018,Sutherland2022}.  The usual anomalous-heating observable is a single diagonal element of a larger object: the spatial electric-field cross spectrum seen by all ions in the device.  Multi-ion probes and collective motional modes are already known to reveal the spatial correlation length and orientation of surface fluctuators \cite{Brownnutt2015,Galve2017}, while scanning-ion measurements can spatially map surface fields and noise \cite{Sagesser2024}.  The question here is different.  We hold the stochastic source statistics fixed and ask how a passive conducting boundary reshapes that complete spatial noise operator.

Patch-potential models naturally separate source statistics from electrostatic propagation through a Green function \cite{Dubessy2009,Low2011,Lin2016}; geometry-resolved calculations have also been used to identify heating hot spots in realistic electrode layouts \cite{Chu2025}.  General surface-induced quantum master equations for charged particles near material interfaces are established \cite{Martinetz2022}.  Our contribution is not the existence of a Green-function heating law or a Lindblad description.  We isolate a geometry-explicit return spectrum and show that it controls the full ion-array covariance, obeys a matrix ordering theorem, and continues beyond the strip through specular saddles of the screened boundary operator.

The parallel slab is the analytically controlled case.  Its image construction is exactly the unfolding of a parallel-wall billiard \cite{Tabachnikov2005}.  The unfolded normal depths form a discrete return measure whose Laplace transform is the electric-field response.  Pairing that measure with the surface spatial spectrum gives the electric-field covariance.  The same construction yields exact special-function heating ratios and a Loewner ordering of the entire many-ion covariance matrix.  Diagonalizing that covariance gives collective environmental noise eigenchannels and an effective noise rank; within a degenerate motional frequency block the corresponding rate matrix yields Lindblad jump operators.  This recasts anomalous heating as one observable of a geometry-dependent correlated bosonic environment.

To distinguish the return picture from a relabeling of the method of images, we test it beyond parallel walls.  Multiple-reflection expansions and stationary-path descriptions of Green functions are classical tools \cite{BalianBloch1970,BalianBloch1972,BalianDuplantier1977,Bordag2001,Zelditch2004}; specular stationary paths themselves are not our novelty.  We use a screened boundary formulation to test whether those paths quantitatively organize the non-flat electrostatic response.  They do: Euclidean path action fixes the exponential decay, boundary curvature fixes the focusing prefactor, and competing saddles add with the expected Laplace weights.

The main results are: (i) an exact return-pair representation of surface-induced electric-field noise; (ii) an $N$-ion covariance theorem showing opposite Loewner ordering for normal and tangential fields under a passive cover; (iii) collective covariance eigenchannels, their blockwise Kossakowski interpretation, and a geometry-dependent effective noise rank; (iv) exact $\zeta(3)$ and $\eta(3)$ single-ion limits, finite-correlation and mode-heating consequences, plus a control-mediated gate-exposure example; and (v) a non-flat screened-BEM test in which billiard saddles predict the electrostatic correction across a controlled geometry family.  The conceptual chain is
\begin{equation}
\begin{aligned}
\text{surface statistics}&\to\text{Green operator}\\
&\to\text{return spectrum}\to C_{ij}(\omega)\\
&\to\text{collective noise channel}.
\end{aligned}
\end{equation}
\section{Surface covariance and the measured heating rate}
\label{sec:source}

Let $G(\bm r,\bm r')$ be the Dirichlet Green function of the vacuum region $\Omega$, and let $\Gamma_e\subset\partial\Omega$ be the fluctuating electrode surface.  Define the sensitivity of field component $j$ at the ion position $\bm r_0$ to a boundary potential at $\bm r'$ by
\begin{equation}
K_j(\bm r';\bm r_0)=\nabla_{\bm r_0}^{(j)}
\frac{\partial G}{\partial n'}(\bm r_0,\bm r').
\label{eq:Kdef}
\end{equation}
For $\omega>0$ we use the one-sided spectral-density convention standard in ion-heating experiments.  With Fourier components defined on the full time axis, we absorb the conventional factor of two into the positive-frequency cross spectrum and write
\begin{align}
&\langle\delta V(\bm r,\omega)\delta V^*(\bm r',\omega')\rangle\nonumber\\
&\qquad=\pi\delta(\omega-\omega')C_V(\bm r,\bm r';\omega),
\qquad \omega,\omega'>0.
\label{eq:covdef}
\end{align}
Then
\begin{equation}
S_{E_j}(\omega)=\iint_{\Gamma_e} K_j(\bm r)K_j^*(\bm r')
C_V(\bm r,\bm r';\omega)\,\dd A\,\dd A'.
\label{eq:generalPSD}
\end{equation}
The spatially local model used below is the deliberate source model
\begin{equation}
C_V(\bm r,\bm r';\omega)=S_{\rm loc}(\bm r,\omega)
\delta_{\Gamma_e}(\bm r-\bm r'),
\label{eq:localcov}
\end{equation}
where $\delta_{\Gamma_e}$ is the surface delta function.  On a physical two-dimensional electrode surface, $S_{\rm loc}$ has units of voltage-noise PSD times area.  Equation~\eqref{eq:generalPSD} then reduces to the single-surface integral
\begin{equation}
S_{E_j}(\omega)=\int_{\Gamma_e}|K_j(\bm r)|^2
S_{\rm loc}(\bm r,\omega)\,\dd A.
\label{eq:localforward}
\end{equation}
A finite patch correlation length instead leaves the double integral in Eq.~\eqref{eq:generalPSD}; we return to that case in Sec.~\ref{sec:correlated}.

For a secular mode of frequency $\omega_j$, neglecting micromotion sidebands, the heating rate in the same one-sided convention is \cite{Turchette2000,Brownnutt2015}
\begin{equation}
\boxed{\displaystyle
\dot{\bar n}_j=\frac{e^2}{4m\hbar\omega_j}S_{E_j}(\omega_j).}
\label{eq:heatingrate}
\end{equation}
Thus every geometric result for $S_{E_j}$ is directly a result for anomalous motional heating.

\section{Many-ion covariance and open-system generator}
\label{sec:opensystem}

For $N$ ions at positions $\bm r_i$ and a chosen motional polarization $u$, define the field cross-spectral matrix
\begin{align}
C^{(u)}_{ij}(\omega)=\iint_{\Gamma_e}&K_u(\bm r;\bm r_i)K_u^*(\bm r';\bm r_j)\nonumber\\
&\times C_V(\bm r,\bm r';\omega)\,\dd A\,\dd A'.
\label{eq:Cijgeneral}
\end{align}
For every complex vector $\bm c$, $\bm c^\dagger C^{(u)}\bm c$ is the spectral density of the collective field $\sum_i c_iE_u(\bm r_i)$ and is nonnegative.  Thus $C^{(u)}(\omega)$ is Hermitian positive semidefinite.  Equation~\eqref{eq:heatingrate} is its $N=1$ diagonal limit; for a normalized mechanical mode vector $\bm b_\nu$, the corresponding field spectrum is
\begin{equation}
S_{E,\nu}(\omega)=\bm b_\nu^\dagger C^{(u)}(\omega)\bm b_\nu,
\qquad
\dot{\bar n}_\nu=\frac{e^2}{4m\hbar\omega_\nu}S_{E,\nu}(\omega_\nu).
\label{eq:modeheatingmatrix}
\end{equation}
This is the standard common/stretch-mode statement generalized to an arbitrary ion array \cite{Brownnutt2015,Galve2017}.

The Green-function-to-master-equation step is established in surface-induced decoherence theory \cite{Martinetz2022}; we use it here to expose the operator consequence of the return geometry.  To make the signed-frequency structure explicit, let
\begin{equation}
\mathcal S_{ij}(\omega)=\int_{-\infty}^{\infty}\dd t\,e^{i\omega t}
\langle E_u(\bm r_i,t)E_u(\bm r_j,0)\rangle
\end{equation}
be the two-sided unsymmetrized quantum spectrum.  For nearly degenerate local oscillators of frequency $\omega_m$, with $H_I=-e x_{\rm zpf}\sum_i(a_i+a_i^\dagger)E_u(\bm r_i)$ and $x_{\rm zpf}=\sqrt{\hbar/(2m\omega_m)}$, the Born--Markov secular generator contains
\begin{align}
\dot\rho={}&-i[H_{\rm eff},\rho]
+\sum_{ij}\Gamma^\downarrow_{ij}\left(a_j\rho a_i^\dagger-\frac12\{a_i^\dagger a_j,\rho\}\right)\nonumber\\
&+\sum_{ij}\Gamma^\uparrow_{ij}\left(a_j^\dagger\rho a_i-\frac12\{a_i a_j^\dagger,\rho\}\right),
\label{eq:kossakowski}
\end{align}
where
\begin{equation}
\Gamma^\downarrow_{ij}=\frac{e^2}{2m\hbar\omega_m}\mathcal S_{ij}(+\omega_m),
\qquad
\Gamma^\uparrow_{ij}=\frac{e^2}{2m\hbar\omega_m}\mathcal S_{ij}(-\omega_m).
\label{eq:ratesigned}
\end{equation}
For classical noise, the one-sided positive-frequency matrix used elsewhere in this paper is twice the corresponding two-sided spectrum, recovering Eq.~\eqref{eq:modeheatingmatrix}.  For quantum surface fluctuations the unsymmetrized spectra at $+\omega$ and $-\omega$ need not coincide; the same geometric construction applies separately to each positive-semidefinite signed-frequency matrix, so upward and downward channel spectra can differ.  For nondegenerate mechanical frequencies the secular approximation is made in the normal-mode basis and Eq.~\eqref{eq:kossakowski} applies within each frequency block; we do not assume that widely separated modes share one Kossakowski matrix.

Within one degenerate or nearly degenerate frequency block, diagonalizing the heating matrix,
\begin{equation}
\Gamma^\uparrow=V\Lambda V^\dagger,
\qquad
L_\mu^\uparrow=\sqrt{\lambda_\mu}\sum_iV^*_{i\mu}a_i^\dagger,
\label{eq:jumps}
\end{equation}
identifies the collective Lindblad jump operators.  More generally, diagonalizing $C^{(u)}(\omega)$ identifies equal-frequency environmental covariance eigenchannels even when distinct mechanical frequencies must be secularized separately.  A scale-independent measure of the number of appreciably populated covariance channels is the participation rank
\begin{equation}
r_{\rm eff}(C)=\frac{(\operatorname{Tr}C)^2}{\operatorname{Tr}(C^2)},
\qquad 1\le r_{\rm eff}\le N.
\label{eq:reffC}
\end{equation}
The return geometry below determines both the eigenvalues and the spatial profiles of these channels.

\section{Parallel slab: exact billiard return measures}
\label{sec:slab}

Consider a noisy plane at $y=0$, an ion at $y=d$, and a passive grounded plane at $y=h>d$.  Let $D$ denote the tangential dimension of the noisy boundary: $D=1$ for a two-dimensional strip cross-section and $D=2$ for a physical three-dimensional slab.

For the normal field, separation parallel to the planes gives
\begin{equation}
\Khat_{\perp,h}(k)=k\frac{\cosh[(h-d)k]}{\sinh(hk)},
\qquad k=|\kvec|,
\label{eq:Kperpclosed}
\end{equation}
while the open half-space has $\Khat_{\perp,\infty}=ke^{-dk}$.  The parallel-plate image series itself is standard in trapped-ion patch-noise theory \cite{Dubessy2009,Brownnutt2015}; here we reorganize it as a geometric return measure.  Unfolding the two parallel boundaries produces the normal depths
\begin{equation}
\alpha_n=|d-2nh|,\qquad n\in\mathbb Z.
\label{eq:alpha}
\end{equation}

\begin{theorem}[Normal return-depth representation]
For every $h>d$ and $k>0$,
\begin{equation}
\boxed{\displaystyle
\Khat_{\perp,h}(k)=k\sum_{n\in\mathbb Z}e^{-\alpha_n k}.}
\label{eq:pathlaplace}
\end{equation}
If $\mu_h^\perp=\sum_n\delta_{\alpha_n}$, then $\Khat_{\perp,h}(k)/k=\mathcal L\{\mu_h^\perp\}(k)$.
\end{theorem}

\begin{figure*}[tbp]
\includegraphics[width=0.94\textwidth]{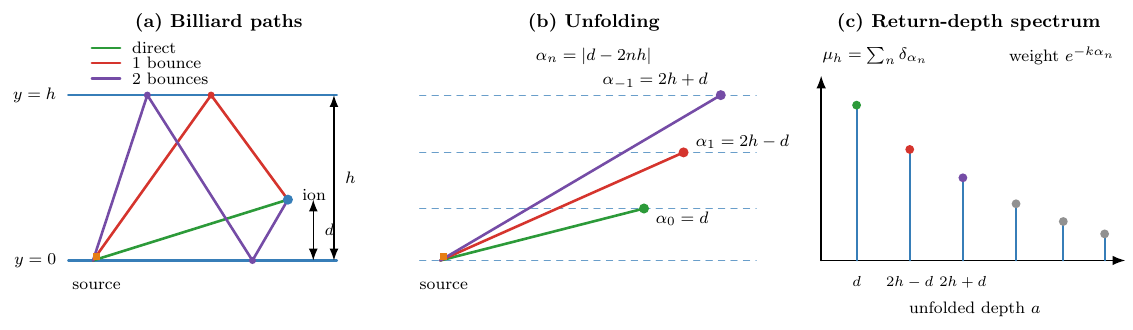}
\caption{Parallel-wall geometry in three equivalent forms.  Left: direct and reflected billiard paths from a point on the noisy plane to the ion.  Center: unfolding turns the reflections into straight paths to image ions.  Right: the corresponding normal-depth measure $\mu_h^\perp=\sum_n\delta_{\alpha_n}$.  Spatial wavenumber Laplace-weights the returns by $e^{-k\alpha_n}$.}
\label{fig:unfolding}
\end{figure*}

The proof is the geometric series obtained from Eq.~\eqref{eq:Kperpclosed}.  Figure~\ref{fig:unfolding} shows the same object as reflected trajectories, unfolded straight paths, and a return-depth spectrum.  The exact transform is in \emph{unfolded normal depth}; it is not the Euclidean length of a laterally displaced physical trajectory.  Euclidean path length appears later as the natural saddle action after the boundary is curved.

Tangential fields carry a different reflection parity.  The potential multiplier at the ion is
\begin{align}
P_h(k)&=\frac{\sinh[(h-d)k]}{\sinh(hk)}\nonumber\\
&=\sum_{r\ge0}e^{-(d+2rh)k}
-\sum_{r\ge0}e^{-(2h-d+2rh)k},
\label{eq:tangentmeasure}
\end{align}
and $\Khat_x=i k_xP_h$.  The two image families therefore enter the tangential response with opposite signs.  Billiard paths are the same geometrically; the field component attaches a parity weight to them.

\section{Anomalous heating as a return-pair functional}
\label{sec:heatingzeta}

Assume a translationally stationary and isotropic source with tangential spectrum $P(k,\omega)$.  Writing $\delta V_{\kvec}(\omega)$ for the tangential Fourier component,
\begin{align}
\langle \delta V_{\kvec}(\omega)\delta V_{\kvec'}^*(\omega')\rangle
&=\pi(2\pi)^D\delta^{(D)}(\kvec-\kvec')\nonumber\\
&\quad\times\delta(\omega-\omega')P(k,\omega),
\qquad \omega,\omega'>0.
\end{align}
For the normal mode, Eqs.~\eqref{eq:pathlaplace} and \eqref{eq:generalPSD} give
\begin{equation}
S_{E_\perp}(\omega)=\int\frac{\dd^Dk}{(2\pi)^D}
 k^2\left(\sum_ne^{-\alpha_n k}\right)^2P(k,\omega).
\label{eq:SEk}
\end{equation}
Define
\begin{equation}
\Phi_\omega(s)=\frac{\Omega_{D-1}}{(2\pi)^D}
\int_0^\infty k^{D+1}P(k,\omega)e^{-sk}\,\dd k.
\label{eq:Phi}
\end{equation}

\begin{proposition}[Return-pair heating functional]
For an isotropic stationary source in the slab,
\begin{equation}
\boxed{\displaystyle
S_{E_\perp}(\omega)=\sum_{m,n\in\mathbb Z}
\Phi_\omega(\alpha_m+\alpha_n).}
\label{eq:returnpairfunctional}
\end{equation}
\end{proposition}

The microscopic source mechanism and the enclosure geometry are cleanly separated: $P(k,\omega)$ determines the scalar transform $\Phi_\omega$, while the billiard supplies the pair-depth set $\alpha_m+\alpha_n$.  This formulation includes finite correlation length without changing the geometric object.

For the local source model, $P(k,\omega)=S_0(\omega)$, and
\begin{align}
S_{E_\perp}&=S_0 C_D\sum_{m,n}(\alpha_m+\alpha_n)^{-(D+2)},\nonumber\\
C_D&=\frac{\Omega_{D-1}\Gamma(D+2)}{(2\pi)^D}.
\label{eq:returnzeta}
\end{align}
Thus local anomalous heating is an inverse-power moment of the billiard pair spectrum.  In the physical $D=2$ slab,
\begin{equation}
\boxed{\displaystyle
S_{E_y}=S_0\frac{3}{\pi}\sum_{m,n}(\alpha_m+\alpha_n)^{-4}.}
\label{eq:physicalzeta}
\end{equation}
For the open half-space this gives
\begin{equation}
S_{E_y,\infty}=\frac{3S_0}{16\pi d^4},\qquad
S_{E_x,\infty}=\frac{3S_0}{32\pi d^4},
\label{eq:openprefactors}
\end{equation}
for either tangential component.  This explicitly recovers the standard local-patch $d^{-4}$ law and the planar normal-to-tangential factor of two \cite{Dubessy2009,Low2011}.  The absolute prefactor is tied to the one-sided surface-PSD normalization in Sec.~\ref{sec:source}; all enclosure ratios below are convention independent.

\subsection{Exact \texorpdfstring{$h=2d$}{h=2d} identities}

At $h=2d$, the positive normal depths are exactly $(2r+1)d$, each once.  The number of ordered odd-positive pairs summing to $2md$ is $m$.  Therefore
\begin{equation}
\sum_{m,n}(\alpha_m+\alpha_n)^{-z}
=(2d)^{-z}\zeta(z-1).
\end{equation}
For the tangential field, the same depths alternate in sign, so the pair multiplicity is $m(-1)^{m-1}$ and the corresponding sum is proportional to the Dirichlet eta function $\eta(s)=(1-2^{1-s})\zeta(s)$.

\begin{corollary}[Zeta/eta heating point]
For local noise at $h=2d$,
\begin{align}
\frac{S_{E_\perp}(2d)}{S_{E_\perp}(\infty)}&=\zeta(D+1),\label{eq:zetageneral}\\
\frac{S_{E_x}(2d)}{S_{E_x}(\infty)}&=\eta(D+1).
\label{eq:etageneral}
\end{align}
For the physical $D=2$ boundary,
\begin{align}
\frac{S_{E_y}(2d)}{S_{E_y}(\infty)}&=\zeta(3)=1.2020569\ldots,\nonumber\\
\frac{S_{E_x}(2d)}{S_{E_x}(\infty)}&=\eta(3)=0.9015427\ldots.
\label{eq:zetaeta3}
\end{align}
At a common analysis frequency, consequently $S_{E_y}/S_{E_x}=8/3$, compared with the open local-patch value $2$.  The two enclosure/open ratios in Eq.~\eqref{eq:zetaeta3} remain valid separately when the normal and tangential secular frequencies differ, provided each frequency is held fixed between the two geometries.
\end{corollary}

The numerical values $1.2021$ and $0.9015$ are therefore exact special-function values, not fitted enclosure gains.  In heating-rate language, Eq.~\eqref{eq:heatingrate} gives the same ratios when the compared mode frequency is held fixed.

\subsection{Mode-selective enclosure theorem}

The zeta/eta point is a special case of a stronger pointwise result.  Relative to the open half-space,
\begin{align}
A_\perp(k)&=\frac{1+e^{-2(h-d)k}}{1-e^{-2hk}}>1,\label{eq:Aperp}\\
A_\parallel(k)&=\frac{1-e^{-2(h-d)k}}{1-e^{-2hk}}<1.
\label{eq:Aparallel}
\end{align}

\begin{theorem}[Mode-selective passive enclosure]
If the stationary source spectrum $P(\kvec,\omega)\ge0$ is unchanged when the passive parallel cover is introduced, then for every $h>d$,
\begin{equation}
S_{E_y,h}(\omega)>S_{E_y,\infty}(\omega),\qquad
S_{E_x,h}(\omega)<S_{E_x,\infty}(\omega),
\label{eq:modeselective}
\end{equation}
for either tangential component $x$ (or $z$) whenever that component has nonzero source weight.
\end{theorem}

\begin{figure}[tbp]
\includegraphics[width=\columnwidth]{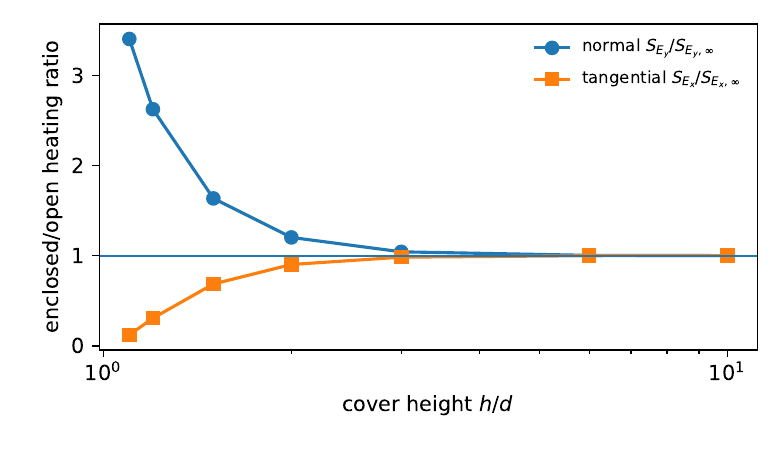}
\caption{Mode-selective anomalous heating for a local source in the physical slab.  A passive cover increases normal heating while suppressing tangential heating.  At $h/d=2$ the exact ratios are $\zeta(3)$ and $\eta(3)$.}
\label{fig:anisotropy}
\end{figure}

The result does not require spatially white patches.  It follows by inserting Eqs.~\eqref{eq:Aperp}--\eqref{eq:Aparallel} under the nonnegative spectral integral.  Reflection parity therefore turns a passive cover into a mode-selective filter: same-sign returns reinforce the normal field, while alternating-sign returns cancel the tangential field.  Figure~\ref{fig:anisotropy} shows the local-source ratios.

\section{Spatial transfer: absolute response versus fractional contrast}
\label{sec:transfer}

Now let the local surface-noise PSD density itself vary in space, $s(\bm x)$.  Translational invariance gives
\begin{equation}
S(\bm x_0)=\int_{\R^D}H_h(\bm x-\bm x_0)s(\bm x)\,\dd^D x,
\qquad H_h=K_{\perp,h}^2.
\label{eq:conv}
\end{equation}
With the Fourier convention $\widehat f(\kvec)=\int e^{-i\kvec\cdot\bm x}f(\bm x)\dd^Dx$,
\begin{equation}
\Hhat_h(\qvec)=\frac{1}{(2\pi)^D}\int
\Khat_h(\bm p)\Khat_h(\qvec-\bm p)\,\dd^Dp.
\label{eq:Hconv}
\end{equation}
Define the total local-source coupling $\G_h=\Hhat_h(0)$ and normalized modulation transfer $M_h(q)=\Hhat_h(q)/\G_h$.

Substituting the return measure gives
\begin{equation}
\Hhat_h(\qvec)=\sum_{m,n}W^{(D)}_{\alpha_m,\alpha_n}(\qvec),
\end{equation}
where
\begin{equation}
W^{(D)}_{a,b}(\qvec)=\frac{1}{(2\pi)^D}\int
|\bm p|\,|\qvec-\bm p|e^{-a|\bm p|-b|\qvec-\bm p|}\,\dd^Dp.
\label{eq:W}
\end{equation}
Every $W^{(D)}_{a,b}$ has a strictly nonnegative integrand.

\begin{proposition}[Absolute modulation response]
For every finite $h>d$ and every finite $q$,
\begin{equation}
\boxed{\displaystyle
\Hhat_h(q)>\Hhat_\infty(q)>0.}
\label{eq:absoluteineq}
\end{equation}
\end{proposition}

The direct pair $(m,n)=(0,0)$ is exactly the open response, and all other path pairs add positively.  This point changes the physical language.  A passive cover does \emph{not} attenuate the absolute modulation amplitude of a local-PSD perturbation.  It adds response, with most of the added weight at coarse spatial scales.  The ``filtering'' effect concerns fractional contrast after division by the increased mean signal.

For
\begin{equation}
s(\bm x)=s_0[1+\epsilon\cos(\qvec\cdot\bm x)],
\end{equation}
we have
\begin{equation}
S(\bm x_0)=s_0\G_h+s_0\epsilon\Hhat_h(q)\cos(\qvec\cdot\bm x_0).
\label{eq:responseabs}
\end{equation}
Thus a detector with fixed additive readout noise benefits from the larger absolute $\Hhat_h$.  The normalized $M_h$ is the appropriate contrast metric when uncertainty or dynamic range scales with the mean response, or when comparing fractional modulation independently of total gain.  Without a metrology noise model, we therefore describe a \emph{gain--contrast redistribution}, not a universal sensitivity tradeoff.

\begin{figure}[tbp]
\includegraphics[width=\columnwidth]{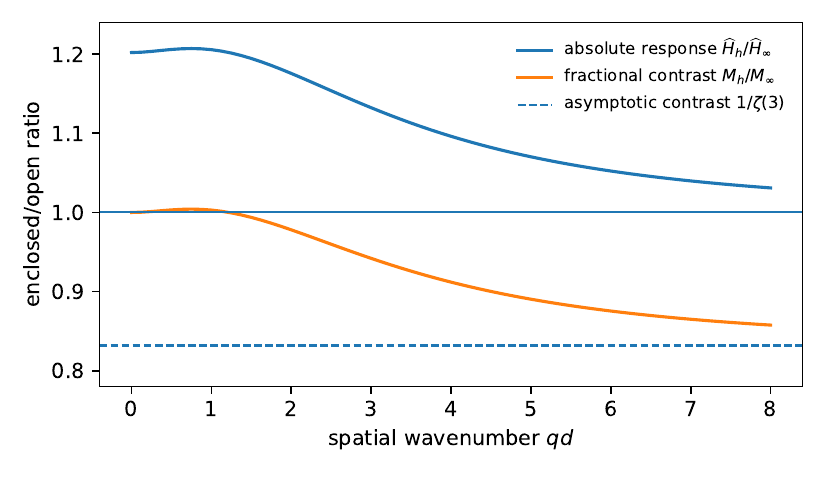}
\caption{Absolute response and fractional contrast at $h/d=2$ for the physical $D=2$ slab.  The enclosure increases $\Hhat$ at every wavenumber, while the normalized response $M_h/M_\infty$ falls below unity over the moderate- and high-$q$ regime and tends to $1/\zeta(3)$.}
\label{fig:gaincontrast}
\end{figure}

Figure~\ref{fig:gaincontrast} makes the distinction explicit for the physical $D=2$ slab at $h/d=2$.  At $qd=3$, the absolute response is $13.23\%$ larger than open while the fractional contrast is $5.80\%$ smaller.  At $qd=4$, the corresponding numbers are $9.62\%$ and $8.80\%$.  The asymptotic normalized ratio $1/\zeta(3)=0.8319\ldots$ is reached only at substantially larger wavenumber.

Because Eq.~\eqref{eq:absoluteineq} gives $\Hhat_h>0$, $M_h(q)>0$ in this geometry.  There is no $\pi$ phase reversal of the local-PSD modulation transfer.

\section{Sharp fine-scale asymptotics and the close-cover crossover}
\label{sec:highq}

The field-level reflected correction is exponentially small for fixed $h>d$:
\begin{equation}
A_\perp(k)-1=\frac{e^{-2(h-d)k}+e^{-2hk}}{1-e^{-2hk}}
\sim e^{-2(h-d)k}.
\label{eq:fieldcrossover}
\end{equation}
The characteristic field crossover is therefore
\begin{equation}
q_\ast\sim\frac{1}{2(h-d)},\qquad
q_\ast d\sim\frac{1}{2(h/d-1)}.
\label{eq:qstar}
\end{equation}
For example, demanding the leading reflected field factor be below $0.1$ requires $qd\gtrsim\ln(10)/[2(h/d-1)]$.  This is $qd\gtrsim2.30$ at $h/d=1.5$ but $qd\gtrsim11.5$ at $h/d=1.1$.  The high-$q$ limit is therefore strongly nonuniform as the cover approaches the ion.  At the formal endpoint $h=d$, where the ion would lie on the grounded boundary and the physical setup ceases to apply, $A_\perp(k)=2/(1-e^{-2dk})\to2$ rather than $1$.

For the heating-transfer convolution in the physical $D=2$ boundary plane, the leading correction is algebraic relative to the direct/direct term.  Let $d=1$ temporarily,
\begin{equation}
f(\kvec)=|\kvec|e^{-|\kvec|},\qquad r_h=\Khat_h-f>0.
\end{equation}
The open transform has the exact Bessel form
\begin{equation}
M_\infty^{(2)}(Q)=\left(\frac{Q^2}{2}+\frac{Q^4}{16}\right)K_0(Q)
+\left(Q+\frac{Q^3}{6}\right)K_1(Q),
\label{eq:Mopen}
\end{equation}
so $f*f\sim c_0Q^{7/2}e^{-Q}$ with
\begin{equation}
c_0=\frac{3\pi}{64}\sqrt{\frac{\pi}{2}}.
\end{equation}
On the other hand, after writing $\qvec=(Q,0)$ and translating the convolution endpoint,
\begin{equation}
\frac{(f*r_h)(\qvec)}{Qe^{-Q}}
\longrightarrow C_h\equiv\int_{\R^2}e^{p_x}r_h(\bm p)\,\dd^2p>0.
\label{eq:crosslimit}
\end{equation}
The positivity of $r_h$ makes the leading coefficient nonzero.

\begin{theorem}[Sharp physical-slab direct-path asymptotics]
For every fixed $d>0$ and $h>d$,
\begin{equation}
\boxed{\displaystyle
\frac{\Hhat_h^{(D=2)}(Q)}{\Hhat_\infty^{(D=2)}(Q)}
=1+c(h/d)(dQ)^{-5/2}+o[(dQ)^{-5/2}],}
\label{eq:sharpasym}
\end{equation}
where $c(\lambda)>0$.  In dimensionless variables,
\begin{equation}
c(\lambda)=\frac{128}{3\pi\sqrt{\pi/2}}
\int_{\R^2}e^{p_x}r_\lambda(\bm p)\,\dd^2p.
\label{eq:ccoef}
\end{equation}
The expansion is uniform on every family $h/d\ge1+\delta$ with fixed $\delta>0$, but not as $\delta\downarrow0$.
\end{theorem}

A proof is given in Appendix~\ref{app:sharp}.  Numerically $c(2)=9.3352$.  The nonuniformity can be quantified: if $h/d=1+\varepsilon$ with $\varepsilon\downarrow0$, then
\begin{equation}
c(1+\varepsilon)\sim 8\sqrt{\frac{2}{\pi}}\,\varepsilon^{-5/2}.
\label{eq:cclose}
\end{equation}
Thus the asymptotic coefficient diverges precisely as the field crossover in Eq.~\eqref{eq:qstar} is pushed to larger $qd$.  The theorem sharpens the earlier $O[(qd)^{-5/2}]$ estimate: the exponent is attained, and the correction approaches zero from above, consistent with Proposition~1.

\section{Finite patch correlations and distance scaling}
\label{sec:correlated}

\begin{figure}[tbp]
\includegraphics[width=\columnwidth]{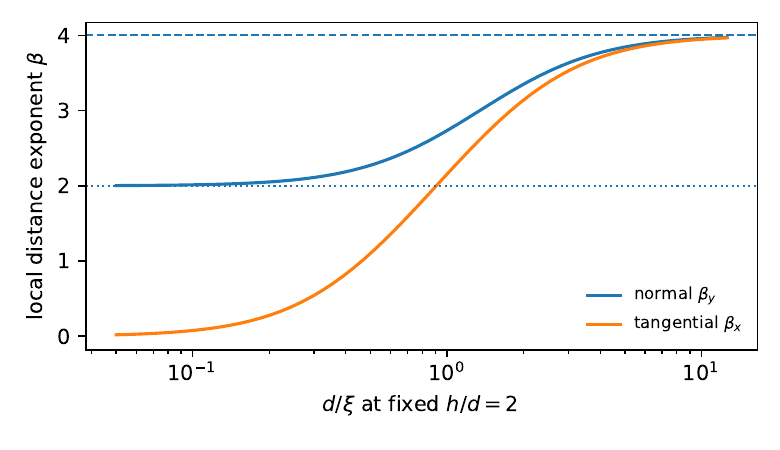}
\caption{Local distance exponents for a Gaussian surface covariance at fixed $h/d=2$.  The normal mode crosses from the capacitor-like $d^{-2}$ regime to the local-patch $d^{-4}$ regime, while the tangential component approaches $d^0$ for very large patches before recovering $d^{-4}$.}
\label{fig:beta}
\end{figure}

Equation~\eqref{eq:returnpairfunctional} already contains a finite correlation length through $P(k,\omega)$.  As an illustrative finite-correlation model, take the Gaussian spatial spectrum
\begin{equation}
P(k,\omega)=A(\omega)e^{-\xi^2k^2/2},
\label{eq:gaussianP}
\end{equation}
we define a local distance exponent while scaling the geometry at fixed $h/d$,
\begin{equation}
\beta_j(d)=-\frac{\partial\ln S_{E_j}}{\partial\ln d}.
\end{equation}
At $h/d=2$, the normal component crosses from $\beta_y\to2$ for $d\ll\xi$ to $\beta_y\to4$ for $d\gg\xi$, reproducing the familiar large-patch to local-patch crossover \cite{Dubessy2009,Brownnutt2015}.  The tangential mode behaves differently: $\beta_x\to0$ in the large-patch limit because a nearly uniform lower-plane potential produces a capacitor-like normal field but no tangential field.  Both components recover $\beta\to4$ for short-range patches.  Figure~\ref{fig:beta} shows the crossover.

The enclosure can also diagnose spatiotemporal source structure.  If
\begin{equation}
P(k,\omega)=A(\omega)B(k),
\end{equation}
then the heating-rate ratio between two geometries at matched secular frequency is independent of $\omega$.  Frequency dependence of that geometry ratio therefore signals a nonseparable source spectrum, for example a correlation length that changes with frequency.  In this sense the enclosure acts as a controlled spatial filter for anomalous-heating mechanism tests.

\section{Many-ion covariance ordering and collective channels}
\label{sec:manyion}

For ions at equal height $d$ with in-plane positions $\bm R_i$, the physical-$D=2$ normal-field matrix is
\begin{equation}
C^{(y)}_{ij,h}(\omega)=\int\frac{\dd^2k}{(2\pi)^2}\,
P(\bm k,\omega)\,|\Khat_{y,h}(\bm k)|^2
 e^{i\bm k\cdot(\bm R_i-\bm R_j)}.
\label{eq:Cijslab}
\end{equation}
The corresponding tangential matrix is obtained by replacing $\Khat_y$ with the chosen tangential component.  Because the slab multipliers in Eqs.~\eqref{eq:Aperp}--\eqref{eq:Aparallel} order the spatial spectrum pointwise, the single-ion theorem upgrades directly to an operator theorem.

\begin{theorem}[Collective covariance ordering]
For any finite equal-height ion configuration and any unchanged stationary source spectrum $P(\bm k,\omega)\ge0$,
\begin{equation}
\boxed{C^{(y)}_h(\omega)-C^{(y)}_\infty(\omega)\succeq0,\qquad
C^{(x)}_\infty(\omega)-C^{(x)}_h(\omega)\succeq0.}
\label{eq:loewner}
\end{equation}
The same tangential statement holds for any in-plane polarization after rotating the coordinates.
\end{theorem}

For example, for the normal component and arbitrary $\bm c$,
\begin{align}
\bm c^\dagger(C_h-C_\infty)\bm c
=\int\frac{\dd^2k}{(2\pi)^2}&P(\bm k,\omega)|\Khat_{y,\infty}|^2
[A_\perp(k)^2-1]\nonumber\\
&\times\left|\sum_i c_i e^{i\bm k\cdot\bm R_i}\right|^2\ge0.
\end{align}
The tangential proof is identical with $1-A_\parallel^2>0$.  Thus every collective normal-field coordinate becomes noisier in absolute terms, even though normalization by the increased diagonal response can make the correlation pattern more concentrated.  Conversely every collective tangential coordinate receives no more absolute noise.  For a stationary quantum source, the same argument applies separately to the positive-semidefinite unsymmetrized spectra $P(\bm k,+\omega)$ and $P(\bm k,-\omega)$, ordering the downward and upward secular rate blocks independently when the underlying source statistics are unchanged.

For an isotropic source the normal covariance depends only on $R_{ij}=|\bm R_i-\bm R_j|$,
\begin{equation}
C^{(y)}_{ij,h}=\frac{1}{2\pi}\int_0^\infty \dd k\,k^3P(k,\omega)
\left(\sum_ne^{-\alpha_n k}\right)^2J_0(kR_{ij}).
\label{eq:CijJ0}
\end{equation}
Equation~\eqref{eq:CijJ0} is the $N$-ion return-pair covariance matrix.  The two-ion result follows by setting $N=2$.  For local noise,
\begin{equation}
S_{12}=\frac{3S_0}{2\pi}\sum_{m,n}
\frac{s_{mn}(2s_{mn}^2-3R^2)}{(s_{mn}^2+R^2)^{7/2}},
\qquad s_{mn}=\alpha_m+\alpha_n.
\label{eq:twoion}
\end{equation}
At $R=2d$, $S_{12}/S_{11}$ changes from $-0.04419$ in the open half-space to $+0.03833$ at $h=2d$.  The spatial common/differential eigenvalue ordering therefore reverses at a common analysis frequency; a reversal of measured COM/stretch heating is not automatic when the two mechanical modes sample different temporal frequencies.

\begin{figure*}[tbp]
\includegraphics[width=0.96\textwidth]{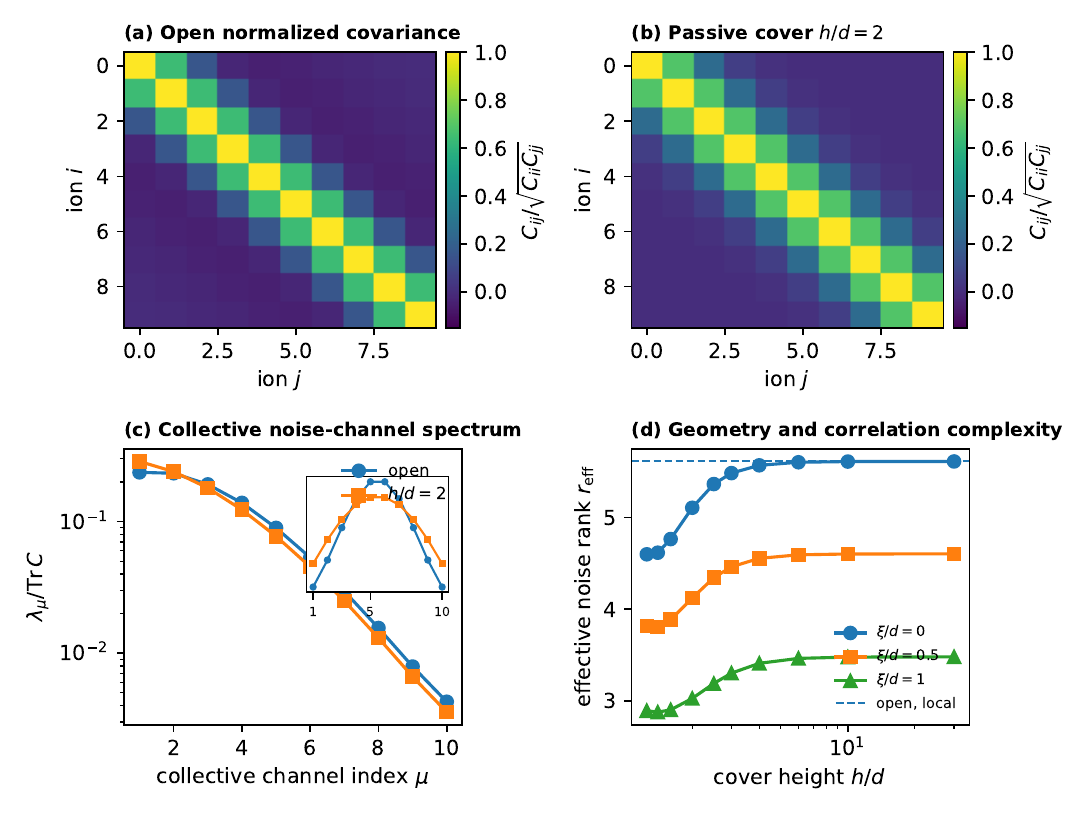}
\caption{Collective normal-field noise for $N=10$ equally spaced probes with spacing $0.6d$.  (a,b) Normalized spatial covariance in the open half-space and at $h/d=2$.  (c) Eigenvalues of the equal-frequency covariance/heating matrix normalized by their sum; the inset shows the leading covariance-eigenchannel profile.  (d) Participation-ratio noise rank versus cover height for local and finite-correlation Gaussian surface spectra.  The cover raises every absolute normal-field quadratic form while concentrating a larger fraction of the total noise into the leading collective channels.}
\label{fig:manyion}
\end{figure*}

Figure~\ref{fig:manyion} shows a ten-ion example with spacing $0.6d$.  For local noise, closing the normal-field cover from the open half-space to $h=2d$ changes $r_{\rm eff}$ from $5.610$ to $5.106$, while the largest normalized eigenvalue increases from $0.2366$ to $0.2863$.  A finite source correlation length concentrates the channel spectrum further: at $\xi=d$ the same ranks are $3.480$ and $3.027$.  The Loewner theorem and the rank reduction are not contradictory: the former orders absolute quadratic forms, whereas the latter describes how the increased total noise is distributed after normalization.

For a uniform microtrap chain the mechanical transverse-mode matrix and the covariance matrix are both nearly translation invariant away from the ends, so their eigenvectors are close to spatial Fourier/cosine modes.  In a ten-site $1/r^3$ transverse-coupling model used only as an illustration, the mean squared overlap between the mechanical normal modes and the environmental eigenvectors is $0.872$ in the open geometry and $0.918$ at $h=2d$; the common-mode share of the total normal-field heating rises from $0.223$ to $0.275$.  For an actual nonuniform Coulomb crystal one should instead project Eq.~\eqref{eq:Cijslab} onto the measured normal-mode eigenvectors using Eq.~\eqref{eq:modeheatingmatrix}.

\subsection{Gate-level heating exposure}
\label{sec:gate}

Motional heating is an established error mechanism for M\o lmer--S\o rensen-type entangling gates, with analytic error models and robust-control strategies already available \cite{Webb2018,Sutherland2022,Kang2023}.  Our purpose here is narrower: to connect the geometry-derived covariance to that established gate-error machinery through the bus-mode projection.  For a gate dominantly using normalized motional vector $\bm b$,
\begin{equation}
\dot{\bar n}_{\rm bus}=\frac{e^2}{4m\hbar\omega_{\rm bus}}
\bm b^\dagger C^{(u)}(\omega_{\rm bus})\bm b.
\label{eq:busheat}
\end{equation}
Any fixed control protocol whose weak-heating error is linear in $\dot{\bar n}_{\rm bus}$ therefore inherits the same enclosed/open ratio.

To make that statement operational, we numerically integrate a primitive single-loop two-ion MS gate with Hamiltonian
\begin{equation}
H(t)=\hbar g(\sigma_x^{(1)}+\sigma_x^{(2)})
(ae^{-i\delta t}+a^\dagger e^{i\delta t}),
\end{equation}
using $g/\delta=1/4$, $t_g=2\pi/\delta$, and an initially ground-state bus.  For the approximately classical electric-field noise modeled elsewhere in this paper, the motional diffusion term is
\begin{equation}
\mathcal L_{\rm diff}\rho=\dot{\bar n}_{\rm bus}
\left(\mathcal D[a]\rho+\mathcal D[a^\dagger]\rho\right),
\end{equation}
so that $\dd\langle n\rangle/\dd t=\dot{\bar n}_{\rm bus}$ in the untruncated oscillator.  Process reconstruction after tracing out the bus gives, for $\dot{\bar n}_{\rm bus}t_g\ll1$,
\begin{equation}
\boxed{\Delta(1-F_{\rm avg})=0.400\,\dot{\bar n}_{\rm bus}t_g
+O[(\dot{\bar n}_{\rm bus}t_g)^2].}
\label{eq:MSheat}
\end{equation}
A Fock-space convergence test gives slopes $0.39984$--$0.39989$ for truncations of seven through ten oscillator levels; details are given in Appendix~\ref{app:gate}.  The coefficient is specific to this primitive gate, whereas the enclosed/open geometry ratio enters only through Eq.~\eqref{eq:busheat}.  At ion separation $R=0.6d$, the $h=2d$ slab multiplies the normal common- and differential-bus heating exposures by $1.237$ and $1.044$, respectively, while tangential exposures are multiplied by $0.878$ and $0.967$.  This is a physically derived bridge from the surface covariance to an established qubit-gate heating budget, not a claim that motional noise is itself a Pauli channel.  The four geometry ratios are summarized compactly as
\begin{center}
\small
\begin{tabular}{lcc}
\toprule
field component & common bus & differential bus\\
\midrule
normal $E_y$ & $1.237$ & $1.044$\\
tangential $E_x$ & $0.878$ & $0.967$\\
\bottomrule
\end{tabular}
\end{center}
for enclosed/open weak-heating exposure at $R=0.6d$ and $h=2d$.

\section{Beyond unfolding: specular saddles of the screened Green operator}
\label{sec:saddles}

Exact unfolding is special to integrable parallel walls.  A more general billiard connection emerges after Fourier transforming a direction $z$ along which the enclosure is invariant.  Write the conserved invariant-direction wavenumber as $q_z$, distinct from the two-dimensional modulation vector $\qvec$ used in Sec.~\ref{sec:transfer}.  The cross-sectional Green function obeys
\begin{equation}
(\nabla_\perp^2-q_z^2)G_{q_z}=-\delta,
\end{equation}
with free kernel
\begin{equation}
G_{q_z}^{(0)}(R)=\frac{1}{2\pi}K_0(q_zR)
\sim\frac{e^{-q_zR}}{\sqrt{8\pi q_zR}}.
\label{eq:K0asym}
\end{equation}
A multiple-reflection expansion expresses boundary corrections as integrals over successive reflection points \cite{BalianBloch1970,Bordag2001}.  The first reflected contribution has the asymptotic form
\begin{align}
I_1(q_z)&=\int_\Gamma A(s,q_z)e^{-q_z\Phi(s)}\,\dd s,\nonumber\\
\Phi(s)&=|\bm r_s-s|+|s-\bm r_0|.
\label{eq:I1}
\end{align}
up to algebraic factors absorbed into $A$.

\begin{proposition}[Specular first-reflection saddle]
Let $s_\star$ be an isolated nondegenerate interior minimum of $\Phi$ on a smooth reflecting boundary, and suppose the reflection amplitude has a regular large-$q_z$ expansion with nonzero leading coefficient at $s_\star$.  Then $s_\star$ satisfies the specular reflection condition, and Laplace's method gives
\begin{equation}
I_1(q_z)\sim A(s_\star,q_z)e^{-q_zL_\star}
\sqrt{\frac{2\pi}{q_z\Phi''(s_\star)}},
\qquad L_\star=\Phi(s_\star),
\label{eq:saddle}
\end{equation}
up to the algebraic $q_z$ dependence already contained in $A$.
\end{proposition}

Indeed, differentiating $\Phi$ along the boundary tangent sets the tangential components of the incoming and outgoing unit vectors equal, which is the mirror-reflection law.  Normal derivatives needed for the electric-field kernel alter the prefactor but not the exponential action.  This is the standard multiple-scattering-to-ray mechanism \cite{BalianBloch1972,BalianDuplantier1977,Zelditch2004}, here applied to the screened electrostatic Fourier operator relevant to surface-noise transfer.

\subsection{Curved-cover numerical test}

We solve the modified-Helmholtz Dirichlet problem with a constant-panel single-layer boundary-element method (BEM) using $K_0(q_zR)/(2\pi)$ as the kernel.  Off-diagonal panel integrals use eight-point Gauss quadrature; the logarithmically singular self-panel integral is evaluated separately by an adaptive change of variables.  The source point is on the noisy lower boundary directly below the ion, and a far cover at $7d$ supplies a numerical direct-response reference.  In the exactly flat $h/d=2$ case, the resulting reflected-response amplitude agrees with the independently integrated strip solution to better than $0.26\%$ over $q_zd=2.5$--$6$.

We fit $\log|K_h/K_{\rm far}-1|$ over the upper half of the sampled range $q_zd=2.5$--$6$ and compare the finite-$q_z$ effective exponent with the independently minimized one-bounce excess length $\Delta L=L_\star-d$.  The production family contains 16 explicitly scripted flat, sinusoidal, phase-shifted, and localized-bump covers.  The Pearson and Spearman correlations between $\Delta L/d$ and the fitted exponent are $0.9991$ and $0.9934$, respectively; the mean absolute error is $0.0204d$ and the median relative error is $0.98\%$.  Restricting to the seven geometries whose saddle is displaced by more than $0.05d$ from the symmetry axis still gives Pearson correlation $0.993$, so the agreement is not supplied only by changing flat-cover height.  After fixing the source collocation point at $x=0$ for every mesh, representative panel-length and finite-difference sweeps give an exponent spread of $4.3\times10^{-5}$ for the flat case and $1.4\times10^{-3}$ for the off-axis depression across target panel lengths $0.14d$--$0.085d$; at fixed $0.11d$ panel length, varying the finite-difference step from $0.007d$ to $0.015d$ changes the off-axis exponent by $2.3\times10^{-4}$.  A separate side-wall sweep over $W/d=4,5,7$ changes the difficult off-axis exponent by only $0.35\%$ and the displaced-wave exponent by $0.012\%$, so the quoted saddle trends are not set by the artificial lateral closure.

\begin{figure*}[tbp]
\includegraphics[width=0.88\textwidth]{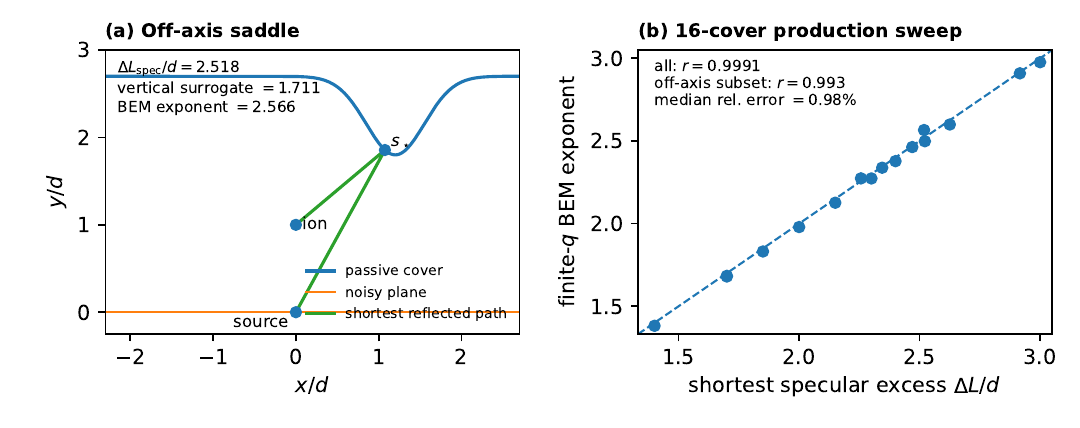}
\caption{Non-flat billiard bridge.  (a) An off-axis cover depression separates Euclidean specular path length from a vertical-depth surrogate; the screened-BEM exponent follows the former.  (b) Across 16 scripted flat and curved covers, the high-$q_z$ reflected-response exponent tracks the independently computed shortest specular excess length with Pearson correlation $0.9991$.}
\label{fig:curved}
\end{figure*}

An off-axis depression separates the competing geometric hypotheses.  Its shortest specular excess length is $\Delta L/d=2.5184$, while a naive vertical-depth candidate is only $1.7115$.  The screened-BEM effective exponent is $2.5658$, selecting Euclidean saddle length rather than projected normal depth by a margin far larger than the discretization spread.  Figure~\ref{fig:curved} shows both this geometry and the 16-shape comparison.

\subsection{Shape derivative, focusing, and multiple saddles}

\begin{figure}[tbp]
\includegraphics[width=\columnwidth]{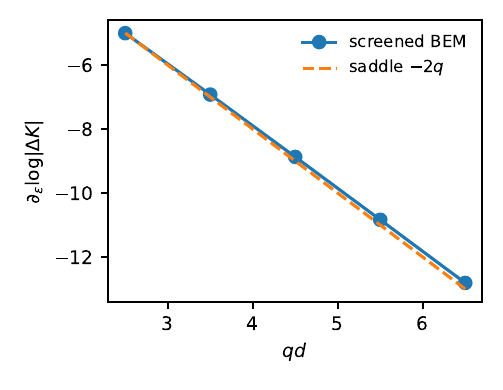}
\caption{Shape derivative of the reflected screened response for a weak sinusoidal cover deformation.  The measured derivative approaches the parameter-free saddle prediction $-2q_z$.}
\label{fig:shape}
\end{figure}

For a centered perturbation $h(x)=h_0+\epsilon\cos(kx)$, the shortest path remains centered for small $\epsilon$ and $\partial_\epsilon\Delta L=2$.  Equation~\eqref{eq:saddle} therefore predicts
\begin{equation}
\partial_\epsilon\log|\Delta K(q_z)|=-2q_z+O(1)
\label{eq:shapepred}
\end{equation}
as $q_z\to\infty$.  A centered BEM difference at $\epsilon/d=\pm0.015$ agrees with the $-2q_z$ coefficient to within $1.50\%$ over $q_zd=2.5$--$6.5$; Fig.~\ref{fig:shape} shows the comparison.  This is the numerical first-order bridge between a boundary deformation and its billiard action.

Path length alone does not determine the amplitude.  We next hold the shortest excess length fixed at $\Delta L=2d$ while varying cover curvature.  At $q_zd=5$, the scaled reflected amplitude $e^{q_z\Delta L}|\Delta K/K_0|$ has correlation $0.9997$ with the Laplace focusing factor $[\Phi''(s_\star)]^{-1/2}$.  After fitting only one overall amplitude scale, the focusing law has a $2.0\%$ rms relative residual and $3.2\%$ maximum residual across the family.  A separate phase sweep creates two competing one-bounce saddles; the scaled BEM amplitude has correlation $0.987$ with the two-saddle predictor $\sum_\gamma e^{-q_z(L_\gamma-L_{\min})}/\sqrt{\Phi''_\gamma}$, with an $8.6\%$ rms relative residual after one overall scale factor.  A two-reflection Laplace test is given in Appendix~\ref{app:saddlenum}.

\begin{figure*}[tbp]
\includegraphics[width=0.88\textwidth]{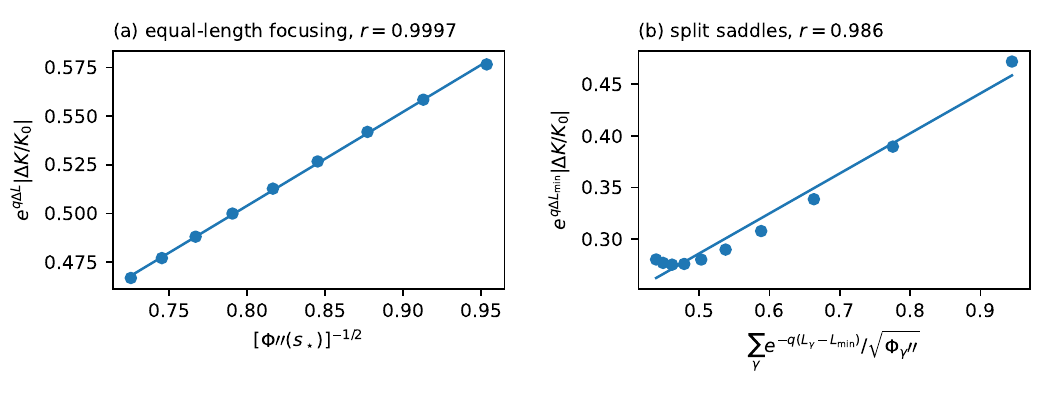}
\caption{Prefactor tests beyond path length alone. (a) Equal-length focusing test at $q_zd=5$: the shortest path length is held fixed while cover curvature changes, and the scaled BEM amplitude follows the Laplace focusing factor with correlation $0.9997$. (b) Two-saddle phase sweep: splitting a pair of short one-bounce saddles changes the BEM amplitude in the direction predicted by the summed exponential/focusing weights, with correlation $0.987$; one-scale residuals are quantified in the text.}
\label{fig:prefactors}
\end{figure*}

The numerical tests support the asymptotic structure
\begin{equation}
\begin{aligned}
\text{boundary reflections}\quad&\longrightarrow\\[-2pt]
&\text{specular saddles at large }q_z.
\end{aligned}
\end{equation}
with exponential action set by path length and amplitude set by focusing and multiplicity.

\subsection{Curved geometry reshapes a collective noise channel}
\label{sec:curvedmulti}

The non-flat saddle calculation and the many-ion channel calculation can be combined without invoking a full three-dimensional trap model.  As a cross-sectional test, we place five equal-height probes at $x/d=(-1.2,-0.6,0,0.6,1.2)$ below a cover with an off-axis depression and form the local-source covariance of the screened $q_zd=0.8$ slice by integrating products of the BEM field kernels over the noisy lower plate.  For the flat cover the leading channel is symmetric and has zero spatial centroid.  The depression shifts its squared-amplitude centroid to $x/d=0.421$ and lowers the participation rank from $2.732$ to $2.578$.  This is not a full physical heating matrix because it is one invariant-direction Fourier slice of a two-dimensional cross-section; its role is to demonstrate that the same curved boundary that selects a specular saddle also reshapes a collective environmental mode.

\begin{figure}[tbp]
\includegraphics[width=\columnwidth]{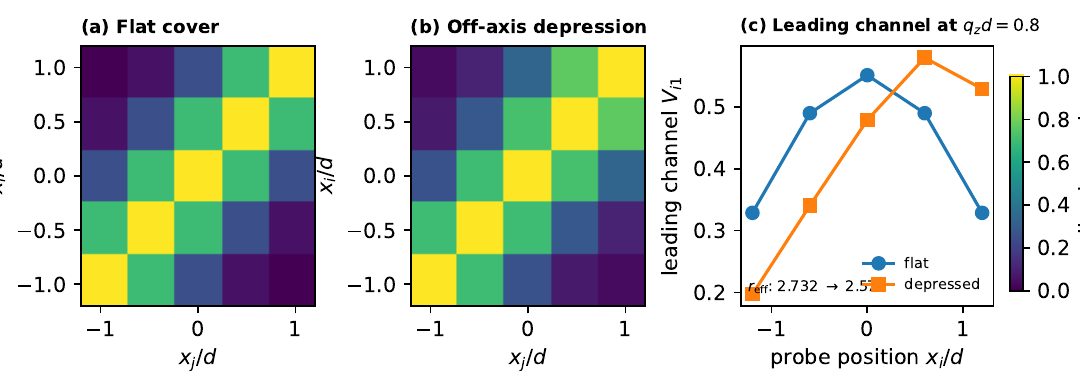}
\caption{Five-probe screened covariance at $q_zd=0.8$ in a cross-sectional model.  An off-axis cover depression breaks the flat-cover symmetry and pulls the leading collective environmental channel toward the depressed side.  This slice test connects the curved-boundary billiard calculation to a correlated multi-probe noise operator.}
\label{fig:curvedmulti}
\end{figure}

\section{Experimental interpretation and scope}
\label{sec:experiment}

A direct test should compare electric-field spectra with and without a passive cover while holding the ion positions, secular frequencies, and mode polarizations fixed by compensating the static trap electrodes.  Three-dimensional scanning-ion measurements already demonstrate the relevant spatial field/noise mapping capability near metallic surfaces \cite{Sagesser2024}.  A simple local-source signature is the exact $h=2d$ anisotropy: normal heating increases by $20.21\%$, tangential heating decreases by $9.85\%$, and at a common analysis frequency the normal-to-tangential spectral ratio increases by $33.3\%$.  A recessed cover, a slotted or transparent electrode pattern, or an imaging axis that does not pass through the cover is required for optical access.

As an illustrative benchmark rather than a device prescription, take $d=50~\mu\mathrm m$, $h=100~\mu\mathrm m$, ion spacing $30~\mu\mathrm m$, and compensated motional frequencies near $2\pi\times1.5~\mathrm{MHz}$.  If the open normal-mode heating is $100$ quanta/s at the matched frequency, the local-source prediction is $120.2$ quanta/s after closing the passive cover.  Under the same source amplitude the open tangential component is half the normal component for the ideal plane and changes from $50$ to $45.1$ quanta/s.  These absolute rates are only a scale choice; the quoted ratios are the geometry predictions.

The added conductor must also be quiet.  At $h=2d$, if it carries independent local noise with PSD density $\chi$ times that on the lower plane, symmetry gives
\begin{equation}
\frac{S_{E_y}}{S_{E_y,\infty}}=(1+\chi)\zeta(3),\qquad
\frac{S_{E_x}}{S_{E_x,\infty}}=(1+\chi)\eta(3),
\end{equation}
so tangential suppression survives only for $\chi<\eta(3)^{-1}-1=0.1092\ldots$.  ``Passive'' is therefore a source-noise requirement as well as an electrostatic boundary condition.

For a simple statistical scale, if one heating-rate estimate has independent $10\%$ relative uncertainty and repeated measurements average as $N^{-1/2}$, five determinations per geometry give roughly $3\sigma$ resolution of the $20.21\%$ normal increase, while about nineteen per geometry are needed for $3\sigma$ resolution of the $9.85\%$ tangential decrease.  Paired interleaving of cover configurations and a ratio measurement can reduce slow drift, but systematic changes in ion position, frequency, polarization, temperature, or surface state are likely to dominate before this ideal counting limit.

The source-side assumption is equally important: the cover changes the transfer operator but not the stochastic process on the original noisy surface.  If it changes adsorbate dynamics, temperature, or patch statistics, geometry and source physics no longer factorize.  The proposed experiment should therefore measure the static trap solution and the heating spectrum at several frequencies in both configurations.  A frequency-dependent geometry ratio would itself be evidence for a nonseparable $P(k,\omega)$ rather than failure of the Green-function framework.

\section{Discussion}

The many-ion formulation changes the interpretation of the slab calculation.  The primary object is not a heating rate but a positive semidefinite spatial noise matrix.  Billiard returns determine its geometric kernel; microscopic surface physics populates spatial wavenumbers through $P(k,\omega)$; and its equal-frequency eigenvectors define collective environmental covariance channels.  Within a degenerate mechanical frequency block, diagonalizing the corresponding Kossakowski matrix gives the physical Lindblad jumps.  Single-ion anomalous heating is one diagonal observable of that channel, while normal-mode heating and gate exposure are quadratic-form projections of the same object.

The Loewner ordering in Eq.~\eqref{eq:loewner} is stronger than a statement about mean heating.  For unchanged source statistics, no collective normal-field coordinate can evade the extra absolute response introduced by a passive parallel cover, whereas every collective tangential coordinate is suppressed.  Yet the normalized channel spectrum can become more concentrated.  The ten-ion example shows exactly this combination: all normal-field quadratic forms increase, while $r_{\rm eff}$ decreases and the leading covariance eigenchannel carries a larger fraction of the trace.  Lower rank should not be equated automatically with easier quantum error correction; it is an experimentally defined property of the motional environment, and a qubit-level error model requires the control-mediated map exemplified by Sec.~\ref{sec:gate}.

The finite-correlation calculation supplies a second control knob.  Increasing the surface correlation length already concentrates the environmental spectrum before the enclosure is added, and the cover shifts it further toward long-wavelength collective coordinates.  This makes geometry a potential diagnostic of spatiotemporal source structure.  If $P(k,\omega)=A(\omega)B(k)$, enclosed/open ratios are frequency independent at matched mode frequency; frequency dependence signals coupled spatial and temporal source structure.

The billiard description also clarifies why the exact slab and the curved-boundary calculations look different.  In the slab, unfolding labels the Green-function terms by normal depths and makes the return-pair heating functional exact.  For a smooth non-flat boundary, the partially Fourier-transformed Green function is a screened boundary problem; specular paths emerge as its large-$q_z$ saddles.  Their Euclidean lengths set exponential actions, their curvatures set focusing determinants, and multiple saddles add coherently at the field-response level.  Integrating and pairing those amplitudes against a source spectrum is what converts pathwise exponential structure into anomalous-heating power laws.

The scope is limited in four ways.  The exact covariance ordering uses an infinite parallel slab and an unchanged stationary source.  The non-flat BEM calculations are cross-sectional models with one invariant direction, not full three-dimensional trap simulations.  The open-system generator assumes the usual Born--Markov/secular regime and must be block-diagonalized in the actual normal-mode basis when mode frequencies are well separated.  The primitive MS calculation is an operational bridge, not a universal gate-error coefficient.  These limitations keep the claims tied to quantities derived from the electrostatics rather than attaching generic QEC language to motional heating.

\section{Conclusion}

Passive conducting geometry can reshape not only the magnitude but the collective structure of electric-field noise seen by trapped ions.  The field cross spectrum $C_{ij}(\omega)$ determines motional heating directly and, within each secular frequency block, the corresponding Kossakowski matrix.  In the parallel slab, its kernel is generated by a billiard return-depth measure, and the same return-pair construction yields exact $\zeta(3)$ and $\eta(3)$ heating factors, finite-correlation scaling, and the full spatial covariance of an ion array.

The collective ordering theorem states that for unchanged source statistics, introducing the passive parallel cover increases the entire normal-field covariance in the Loewner order and decreases the entire tangential-field covariance.  Diagonalization identifies collective equal-frequency noise eigenchannels; in a degenerate block the same construction gives Lindblad jump operators.  In the ten-ion example the cover concentrates a larger fraction of normal noise into the leading channel and lowers the effective rank, while an explicit primitive entangling-gate simulation shows how the projected covariance enters a qubit-gate heating budget.

The billiard connection is not restricted to the integrable strip.  After Fourier transforming an invariant direction, reflected contributions to a smooth non-flat Green operator are controlled at large $q_z$ by specular saddles.  Corrected boundary-element calculations distinguish Euclidean saddle length from projected depth, resolve the focusing prefactor, and track competing paths across a controlled curved-cover family.  A five-probe curved example shows that the same boundary deformation also shifts a collective noise eigenvector.

The resulting chain is
\begin{equation}
\begin{aligned}
\text{surface physics}&\to\text{billiard geometry}\\
&\to C_{ij}(\omega)\to\text{noise eigenchannels}\\
&\to\text{motional/gate exposure}.
\end{aligned}
\end{equation}
It provides a route for using passive geometry both as a diagnostic of anomalous surface noise and as a controlled way to reshape the correlated bosonic environment of trapped-ion quantum hardware.

\begin{acknowledgments}
We thank Prof.~Chris Cox and Prof.~Robert Niffenegger for supervision and helpful discussions during the honors-thesis project from which this work developed.  We also thank Chris Caron and Zhenyu Wei for regular project discussions.
\end{acknowledgments}

\section*{Data and code availability}
Scripts and recorded numerical data for the slab, many-ion covariance matrices, primitive gate simulation, ray reconstruction, screened boundary-element calculations, and billiard-saddle tests are provided with the reproducibility repository at \url{https://github.com/lostree9/noise_recycling}.

\appendix

\section{Sharp \texorpdfstring{$D=2$}{D=2} asymptotics}
\label{app:sharp}

Set $d=1$ and $\qvec=(Q,0)$.  Write $\Khat_h=f+r_h$ with $f(k)=ke^{-k}$ and $r_h>0$.  From Eq.~\eqref{eq:Aperp},
\begin{equation}
r_h(k)=ke^{-k}\frac{e^{-2(h-1)k}+e^{-2hk}}{1-e^{-2hk}}.
\end{equation}
For fixed $h>1$, $e^k r_h\in L^1\cap L^\infty$ with the first radial moment integrable.  Translating the endpoint in the cross convolution gives
\begin{align}
(f*r_h)(Q\hat x)
&=\int f(Q\hat x-\bm p)r_h(\bm p)\,\dd^2p\nonumber\\
&=Qe^{-Q}\left[\int e^{p_x}r_h(\bm p)\dd^2p+o(1)\right],
\end{align}
by dominated convergence.  The integral is strictly positive.

The exact open transfer, Eq.~\eqref{eq:Mopen}, and $K_\nu(Q)\sim\sqrt{\pi/(2Q)}e^{-Q}$ give
\begin{equation}
(f*f)(Q\hat x)\sim \frac{3\pi}{64}\sqrt{\frac{\pi}{2}}Q^{7/2}e^{-Q}.
\end{equation}
The reflected/reflected term is $O(e^{-Q})$ after the same exponential weighting.  Hence
\begin{equation}
\frac{(f+r_h)*(f+r_h)}{f*f}
=1+\frac{2C_h}{c_0}Q^{-5/2}+o(Q^{-5/2}),
\end{equation}
which proves Theorem~3.  If $h/d\ge1+\delta$, the weighted norms of $r_h$ are bounded uniformly for fixed $\delta>0$, giving uniformity on that family.  As $\delta\downarrow0$, the bound diverges and the limit is nonuniform.  More precisely, for $h=1+\varepsilon$ radial symmetry gives
\begin{equation}
C_h=2\pi\int_0^\infty k^2 I_0(k)
 e^{-k}\frac{e^{-2\varepsilon k}+e^{-2(1+\varepsilon)k}}{1-e^{-2(1+\varepsilon)k}}\,\dd k.
\end{equation}
Using $I_0(k)\sim e^k/\sqrt{2\pi k}$ and scaling $k=u/\varepsilon$ yields
\begin{equation}
C_{1+\varepsilon}\sim\frac{3\pi}{16}\varepsilon^{-5/2},
\end{equation}
which inserted into Eq.~\eqref{eq:ccoef} gives Eq.~\eqref{eq:cclose}.  The earlier elementary direct/direct lower-bound proof may use
$|\bm p|+|\qvec-\bm p|\le Q+4$ on $Q/4\le p_x\le3Q/4$, $|p_y|\le\sqrt Q$; the constant is $4$, not $3$.

\section{Tangential signed return measure}

Equation~\eqref{eq:tangentmeasure} defines a signed depth measure
\begin{equation}
\mu_h^\parallel=\sum_{r\ge0}\delta_{d+2rh}-
\sum_{r\ge0}\delta_{2h-d+2rh}.
\end{equation}
For isotropic $P(k,\omega)$, angular averaging gives one tangential component
\begin{equation}
S_{E_x}=\frac{1}{D}\iint\Phi_\omega(a+b)\,
\dd\mu_h^\parallel(a)\dd\mu_h^\parallel(b).
\end{equation}
At $h=2d$, $\mu^\parallel=\sum_{r\ge0}(-1)^r\delta_{(2r+1)d}$, and the ordered-pair coefficient at total depth $2md$ is $m(-1)^{m-1}$, which gives Eq.~\eqref{eq:etageneral}.  The open local-source anisotropy is $S_{E_y}/S_{E_x}=D$; at $h=2d$ it becomes
\begin{equation}
\frac{S_{E_y}}{S_{E_x}}=
D\frac{\zeta(D+1)}{\eta(D+1)}=
\frac{D}{1-2^{-D}}.
\end{equation}

\section{Ray reconstruction identity}
\label{app:ray}

Let $P_a(x)=a/(x^2+a^2)$ and $\sigma_n=\operatorname{sgn}(d-2nh)=\partial_d\alpha_n$.  For a circular gate of radius $\rho$ around the unfolded image at depth $a$, the cosine-measure crossing probability is $\rho P_a(x)$.  Splitting the expected crossing counts by $\sigma_n$ and differentiating in $d$ yields
\begin{equation}
\frac{1}{\rho}\left(\partial_dg^\uparrow-\partial_dg^\downarrow\right)
=\sum_n\frac{x^2-\alpha_n^2}{(x^2+\alpha_n^2)^2}
=-\pi K_h(x).
\end{equation}
A centered height difference produces the Monte Carlo estimator used in Sec.~\ref{sec:ray}.  Common random numbers at $d\pm\delta$ reduce finite-difference variance.

\section{Constructive billiard reconstruction in the strip}
\label{sec:ray}

In the parallel strip, unfolding also provides a constructive estimator for the real-space field kernel.  Launch rays from a boundary point $x$ under the cosine angular measure and count crossings of a circular gate centered on the ion, keeping upward and downward crossings separate.  For one unfolded image at depth $a$, the crossing probability is $\rho a/(x^2+a^2)$.  Differentiating the signed crossing count with respect to ion height reconstructs the image derivative and hence $K_h(x)$.  The derivation is recorded in Appendix~\ref{app:ray}.

This calculation should not be read as an independent validation of image electrostatics: the estimator is derived from the same exact unfolding identity.  Its value is constructive.  It shows that the spatial-transfer observable can be reconstructed from direction-resolved billiard data.  At $h/d=2$, 31 launch points spanning $x/d\in[-4,4]$, 8000 rays per point, five stratified seeds, gate radius $\rho/d=0.429$, and centered height step $\delta/d=0.10$ give Pearson correlation $0.99993$ with the exact field kernel.  The reconstructed transfer half-power point differs from the exact value by $0.78\%$.

\begin{figure*}[tbp]
\centering
\includegraphics[width=0.70\textwidth]{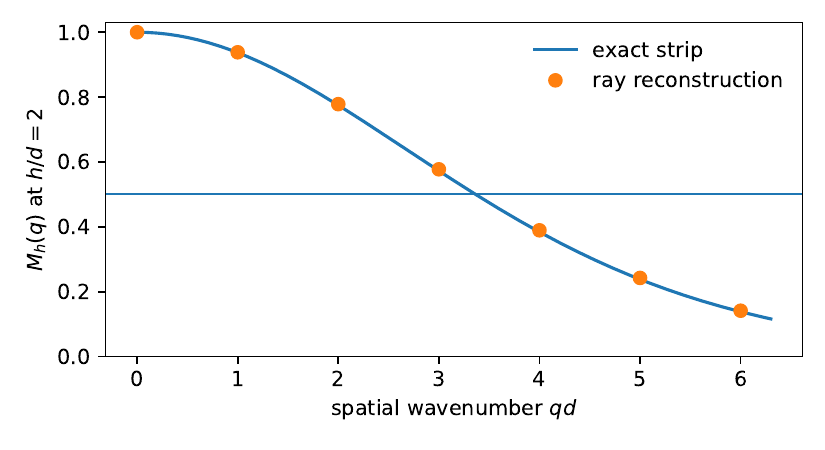}
\caption{Exact and direction-resolved ray reconstruction of the strip modulation transfer at $h/d=2$.  The calculation is a constructive reconstruction from billiard crossings, not an independent electrostatic model.}
\label{fig:ray}
\end{figure*}

\section{Curved-cover screened-BEM details}
\label{app:saddlenum}

The cross-section is closed by remote side walls at $|x|=5d$.  Smooth covers are represented by piecewise-linear panels.  At each longitudinal Fourier wavenumber $q_z$, the single-layer matrix uses the free kernel $K_0(q_zR)/(2\pi)$.  Eight-point Gauss quadrature is used off the diagonal, while the logarithmically singular self-panel integral is evaluated after the substitution $s=(\Delta\ell/2)t^2$ with adaptive quadrature.  Up to the conventional sign of the single-layer representation, the solved boundary density is the boundary normal derivative entering the Dirichlet Green function; a centered displacement of the ion gives the normal field sensitivity at the source point.

For the 16-geometry production sweep, target panel length is $0.11d$, the field finite-difference step is $0.01d$, and the exponential fit uses the upper half of $q_zd=2.5$--$6$.  The flat $h/d=2$ geometry predicts $\Delta L/d=2$ and gives finite-$q_z$ fitted exponent $1.9789$; its full reflected amplitude agrees with the exact partially Fourier-transformed strip result to within $0.26\%$ over the sampled band.  A representative displaced wave predicts $\Delta L/d=2.5218$ and gives $2.4981$.  The lower noisy plate is panelized with a collocation midpoint fixed exactly at $x=0$ at every resolution.  Changing target panel length from $0.14d$ to $0.085d$ changes the fitted exponent by $4.3\times10^{-5}$ for the flat case and $1.4\times10^{-3}$ for the deliberately difficult off-axis depression, whose correction reaches only $\sim4\times10^{-8}$ at the largest sampled $q_z$.  At fixed $0.11d$ panel length, varying the finite-difference step from $0.007d$ to $0.015d$ changes the off-axis exponent by $2.3\times10^{-4}$.  All quoted shortest saddles lie well inside the remote side walls.  We nevertheless repeated two representative calculations at $W/d=4,5,7$.  For the off-axis depression the fitted exponents are $2.5630$, $2.5658$, and $2.5720$, a total spread of $0.0090$ ($0.35\%$ of the $W=5d$ value).  For the displaced wave they are $2.49815$, $2.49807$, and $2.49786$, a $0.012\%$ spread.

To verify the reflection-order Laplace principle directly, we also evaluate scalar one- and two-reflection action integrals for a representative curved cover.  The independently minimized actions are $L_1/d=3.180$ and $L_2/d=5.151$; fits to $\log I_n(q_z)$ over the finite range $q_zd=5.5$--$8$ give slopes $3.258$ and $5.310$, approaching the corresponding minima from above as expected from the algebraic Laplace prefactors.

\section{Primitive gate-channel simulation}
\label{app:gate}

For the calculation in Sec.~\ref{sec:gate}, we propagate each of the 16 qubit operator-basis elements tensored with the oscillator vacuum under the time-dependent Hamiltonian and the symmetric diffusion generator $\dot{\bar n}(\mathcal D[a]+\mathcal D[a^\dagger])$.  The ideal target is $U=\exp(i\pi\sigma_x^{(1)}\sigma_x^{(2)}/4)$, up to a global phase.  After the gate time $t_g=2\pi/\delta$, the oscillator is traced out and the average gate fidelity is obtained from the reconstructed qubit channel.

The quoted weak-heating coefficient is extracted after subtracting the finite-Fock no-noise floor and fitting the three smallest nonzero values $\dot{\bar n}t_g=10^{-4},2\times10^{-4},5\times10^{-4}$.  The fitted slopes for oscillator truncations $n_{\rm ph}=5,6,7,8,9,10$ are $0.40873$, $0.40044$, $0.39989$, $0.39984$, $0.39984$, and $0.39984$, respectively.  We therefore quote $0.400$ rather than attaching significance to the fourth decimal place.

\end{document}